\documentclass[preprint,prl,aps,amsmath,amssymb,color,superscriptaddress,floatfix]{revtex4}
\usepackage{stmaryrd}
\usepackage{amsmath}
\usepackage{amssymb}
\usepackage{graphicx}
\usepackage{dcolumn}
\usepackage{bm}
\usepackage{tabularx}
\usepackage{color}
\usepackage[normalem]{ulem}
\begin{document}
\begin{titlepage}
		\title{First-principles design of excitonic insulators: A review}
		
		\author{Hongwei Qu}
		\affiliation{Key Lab of advanced optoelectronic quantum architecture and measurement (MOE), and School of Interdisciplinary Science, Beijing Institute of Technology, Beijing 100081, China}
		\author{Haitao Liu}
		\affiliation {Institute of Applied Physics and Computational Mathematics, Beijing 100088, China}
		\affiliation {National Key Laboratory of Computational Physics, Beijing 100088, China}
		\author{Yuanchang Li}
		\email{yuancli@bit.edu.cn}
		\affiliation{Key Lab of advanced optoelectronic quantum architecture and measurement (MOE), and School of Interdisciplinary Science, Beijing Institute of Technology, Beijing 100081, China}
		
		\date{\today}
		
\begin{abstract}
The excitonic insulator (EI) is a more than 60-year-old theoretical proposal that yet remains elusive. It is a purely quantum phenomenon involving the spontaneous generation of excitons in quantum mechanics and the spontaneous condensation of excitons in quantum statistics. At this point, the excitons represent the ground state rather than the conventional excited state. Thus, the scarcity of candidate materials is a key factor contributing to the lack of recognized EI to date. In this review, we begin with the birth of EI, presenting the current state of the field and the main challenges it faces. We then focus on recent advances in the discovery and design of EIs based on the first-principles Bethe-Salpeter scheme, in particular the dark-exciton rule guided screening of materials. It not only opens up new avenues for realizing excitonic instability in direct-gap and wide-gap semiconductors, but also leads to the discovery of novel quantum states of matter such as half-EIs and spin-triplet EIs. Finally, we will look ahead to possible research pathways leading to the first recognized EI, both computationally and theoretically.
\end{abstract}
		
		\maketitle
		\draft
		\vspace{2mm}
\end{titlepage}

\vspace{0.3cm}
\textbf{1. INTRODUCTION}
\vspace{0.3cm}

The past two decades have witnessed tremendous achievements in first-principle design of novel quantum materials, especially topological matters. In 2017, Varsano \emph{et al}. employed the first-principles Bethe-Salpeter scheme to predict that carbon nanotubes are intrinsic exciton insulators (EIs)\cite{Varsano2017}. In 2018, Jiang \emph{et al}. proposed a dark-exciton rule for the occurrence of excitonic instability, which provides a useful guideline for finding and designing EIs\cite{Jiang2018}. Since then, researchers have successfully predicted a wide variety of EIs through this approach, including EIs via band engineering, topological EIs, and magnetic EIs. The rise of two-dimensional materials has rekindled the 60-year-old EI mystery, leading to advances in materials, theory, and experiment. In this review, we will focus on advances in first-principles design of EI candidates. For other aspects, see the reviews, e.g., Refs. \onlinecite{Kaneko,Kune,HuangY,Wu2025}.

\vspace{0.3cm}
\textbf{1.1. The birth of EI}
\vspace{0.3cm}

In the 1930s, Frenkel\cite{Frenkel1931a,Frenkel1931b} and Peierls\cite{Peierls} conceptualised excitons as a fundamental excitation of solids, which are electron-hole pairs bound together by Coulomb attraction. On scales larger than the radius of the exciton, they can be regarded as bosons and are therefore capable of Bose-Einstein condensation\cite{Blatt}. For the first 30 years, excitons were studied as excited states. It was not until 1961 that Mott first suggested that in semimetals with low carrier concentrations, excitons might spontaneously arise and open an energy gap, leading to a metal-insulator phase transition\cite{Mott}. Two years later, Knox proposed a condition for spontaneous exciton generation (excitonic instability), i.e., the exciton binding energy $E_b$ exceeds its single-electron excitation gap $E_g$\cite{Knox}. With further developments by Keldysh, Kopaev \cite{Keldysh}, and DesCloizeaux\cite{DesCloizeaux}, J\'{e}rome, Rice, and Kohn proposed that, for semimetals with a small overlapping of valence and conduction bands or for semiconductors with a small bandgap, there may exist, at low temperatures, a new state of matter in which excitons are spontaneously generated and condense to a BCS-like state\cite{Jerome}. They named the matter with such a state as the EI.

Later, Halperin and Rice\cite{Halperin} gave a more precise definition of an EI. Here we quote their original words ``\emph{Let us focus our attention on the crystal at 0 K, in the semiconducting region with $E_g$ close to $E_b$. For $E_g > E_b$, no excitons are present in the ground state of the crystal, and the nondistorted state is stable. For $E_g < E_b$, excitons are present. In this region, treating the crystal in a Hartree-Fock approximation is roughly equivalent to treating the excitons as a weakly repulsive Bose gas. Excitons form until the repulsive potential cancels the negative energy ($E_g - E_b$) associated with the creation of a single exciton. Furthermore, and most important, the excitons present will form a Bose condensate in the exciton state of minimum energy, namely the state with wave vector \textbf{w}.}"

The EI is another macroscopic quantum state analogous to superconductors and superfluids, essentially originating from many-body interactions between electrons. It involves spontaneous generation of excitons in quantum mechanics and spontaneous condensation in quantum statistics. At this point, the exciton is the ground state of the system, not the excited state\cite{Halperin,Kohn}. Now single-electron picture breaks down and the system displays the bosonic physics. The study of EI contributes to a deeper understanding of many-body interactions in condensed-matter physics and helps to shed light on the superconductivity/superfluidity mechanism. Whether or not EI correlates to some kind of supertransport phenomenon could provide more evidence for the study of the relationship between Bose condensation and superfluidity\cite{Jerome,Halperin,Kohn,Zittartz,Hanamurat,Eisenstein,Mazza}. Since exciton binding is typically much stronger than electron Cooper pairs, EIs have a greater potential to maintain quantum correlations at or above room temperature, thus providing a new material platform for quantum information technology.

\vspace{0.3cm}
\textbf{1.2. Developments and bottlenecks}
\vspace{0.3cm}

Despite a new era, EI remains elusive\cite{Kaneko}. The lack of recognized EI to date is mainly due to the following two limitations.

(i) Scarcity of material platforms. Excitons were originally introduced to explain the photoexcitation behaviour of semiconductors, and hence they are used to describe an excited state rather than a ground state. In other words, spontaneous generation of excitons is very rare. Known EI candidates, both those predicted in theory and those with experimental tests, can be broadly classified into three categories. The first category is semimetals with small overlapping between valence and conduction bands, or semiconductors with small bandgaps, which were proposed at the inception of EI. This list includes transition-metal dichalcogenides\cite{Cercellier,Kogar,Rossnagel,GaoNP,GaoNC,Song,YangHNJP,Samaneh}, Ta$_2$NiSe$_5$\cite{Mazza,Wakisaka,npj2021,Guo2024}, Ta$_2$Pd$_3$Te$_5$\cite{HuangJ,ZhangP,YaoJ,Hossain,JiangBei}, and some other nanostructures\cite{LiZ,WangJR,ZhuZW}. Among them, TiSe$_2$ and Ta$_2$NiSe$_5$ are representative star materials that have received intensive attention. They have the advantage of the presence of easily experimentally recognisable phase transition signals, but the disadvantage of being accompanied by structural distortions. Experimentally, there is a lack of robust means to distinguish whether the phase transition is exciton-driven or simply originates from the Jahn-Teller effect\cite{Rossnagel,npj2021}.

The second category is the electron-hole bilayer structure, which is traditionally realized using semiconductor quantum wells\cite{Du,WangR}. The emergence of two-dimensional van der Waals heterojunctions in recent years has rapidly made this paradigm a hot topic in EI researches\cite{Liu,Gu,Chen,Xu,Tan,Lian}. The spatial separation of electrons and holes allows for longer lifetimes of interlayer excitons, and non-equilibrium exciton condensation at $\sim$100 K has been experimentally demonstrated\cite{Wang}. However, the ``exciton" here is strictly ``electron-hole composite fermion pair", which is insufficient to explain all the features of ``real" excitons. Despite some inherent shortcomings of this scheme in fully understanding EI, such as spontaneous symmetry breaking, it does provide a wealth of useful insights. In this review, our focus is on excitonic instability in intrinsic materials

The third category is theoretically suggested wide-gap semiconductors in recent years, mainly in low-dimensional systems\cite{Jiang2018,Jiang2019,Jiang2020,Dong2020,Dong2021,Sethi,Liu2021,Liu2022,Liu2024,Zhao}. It shows that excitonic instabilities can occur even in strongly-correlated semiconductors with $E_g >$ 3 eV\cite{Jiang2019,Jiang2020}. The wide-gap naturally suppresses the Jahn-Teller distortions, thus avoiding the intrusion of structural mechanism. The drawback, however, is that the insulator-insulator phase transition in this case loses an easily recognizable signal of the kind seen experimentally in metal-insulator transition. This raises a new question of how to distinguish between a many-body gap and a single-electron gap\cite{Liu2021}.

(ii) Lack of a deterministic identification method for experiments. As the definition of EI in \textbf{1.1}, it is a theoretical mechanism whichever is not defined by phenomena. This contrasts with superconductors, which are founded on two well-defined experimental phenomena: zero resistance and the Meissner effect. A more practical challenge is that most contemporary exciton detection techniques rely on observing the evolution of systems under external energy injection\cite{Kasprzak}. By contrast, the EI is precisely the ground state and its appearance does not require an input of energy. Although for a given material, spontaneous condensation of excitons can be inferred by comparing the crystal structure, characteristics of the frontier state, and optical properties of the high-temperature and low-temperature phases\cite{Cercellier,Wakisaka,Du,Bucher}. However, at this stage of knowledge, the presence of all these signals can only indicate that the exciton mechanism may be one of the possibilities, and it is not possible to completely rule out other competing mechanisms, such as the Jahn-Teller, the Mott, and so on. This is the fundamental reason why EIs are still full of controversy after more than half a century of research. On the other hand, some researchers have attempted to determine spontaneous exciton condensation by measuring the superflow phenomenon, but the charge-neutrality of excitons makes electrical measurements extremely challenging. In addition, it is theoretically controversial whether EIs are capable of generating some kind of superfluidity since they have a diagonal long-range order\cite{Kohn,Zittartz,Hanamurat,Eisenstein}.

\vspace{0.3cm}
\textbf{2. First-principles Bethe-Salpeter scheme}
\vspace{0.3cm}

As defined by Halperin and Rice\cite{Halperin}, an EI is simply a substance for which $E_g < E_b$ at 0 K. First-principles Bethe-Salpeter scheme is able to quantify these two quantities, which makes it a powerful tool for finding and designing EIs.

To obtain $E_g$ using the methods based on the DFT, one needs to solve the equation
\begin{eqnarray}
	  \{-\frac{1}{2}\nabla^2 + v_{nu}({\bf r}) + v_H([n];{\bf r}) + \Theta\} \phi_i({\bf r}) = \varepsilon_i\phi_i({\bf r}),\ \ [n] = \sum_i^N|\phi_i({\bf r})^2|.
\end{eqnarray}
Here the first three terms refer to the kinetic energy, the nuclear and the Hartree potential, which remain formally consistent regardless of the calculation method. The difference between the different methods is reflected in the approximation to the fourth term $\Theta$. For the standard DFT, it is given by local or semi-local exchange-correlation potentials. However, such approximations systematically underestimates $E_g$ of semiconductors and insulators by more than 30\%\cite{Aryasetiawan,Tanpccp,TanJCP}. To overcome this problem, $\Theta$ in GW employs a non-local and energy-dependent potential (self-energy). However, compared with DFT, the GW greatly increases the computational cost, so a balance between computational accuracy and efficiency must be considered in practice. At present, in most cases, a single-point G$_0$W$_0$ is used instead of a fully self-consistent GW\cite{Aryasetiawan,Jiang2020}.

To obtain $E_b$, one needs to solve the Bethe-Salpeter equation (BSE), which reduces to an eigenvalue problem of the Hamiltonian\cite{Marini}
\begin{eqnarray}
	  H_{vckv'c'k'} = (\varepsilon_{ck} - \varepsilon_{vk}) \delta_{vv'} \delta_{cc'} \delta_{kk'} + (f_{ck} - f_{vk})[2V_{vckv'c'k'}-W_{vckv'c'k'}]
\end{eqnarray}
The first term comes from single-electron eigenvalues of Eq. (1). The second term is the BSE kernel, where $W_{vckv'c'k'}$ and $V_{vckv'c'k'}$ represent an attractive screened electron-hole interaction and a repulsive Coulomb exchange interaction, respectively. Sometimes, the ``scissor operator" is used to deal with the bandgap problem. This method is relatively less computationally demanding and in good agreement with experiment\cite{Tiago,Dong2021}. In view of the fact that different computational methods usually change the band shape slightly, applying a scissor operator $\Delta$ to rigidly shift the conduction band away from the valence band by $\Delta$ yields a new single-electron band, which is then taken as a new input for solving the BSE. Notably, it not only changes the first term of the Hamiltonian (2) from $\varepsilon_{ck} - \varepsilon_{vk}$ to $\varepsilon_{ck} - \varepsilon_{vk} + \Delta$, but also leads to the change of $W_{vckv'c'k'}$, which in turn changes $E_b$\cite{Dong2021}. Noteworthy, the scissor operator is only a compromise option to a self-consistent GW computation, and its reliability needs to be carefully examined.

\vspace{0.3cm}
\textbf{3. First-principles design of EIs}
\vspace{0.3cm}

\vspace{0.3cm}
\textbf{3.1. Dark-exciton rule}
\vspace{0.3cm}

%fig01
\begin{figure}[htbp]
\includegraphics[width=0.95\columnwidth]{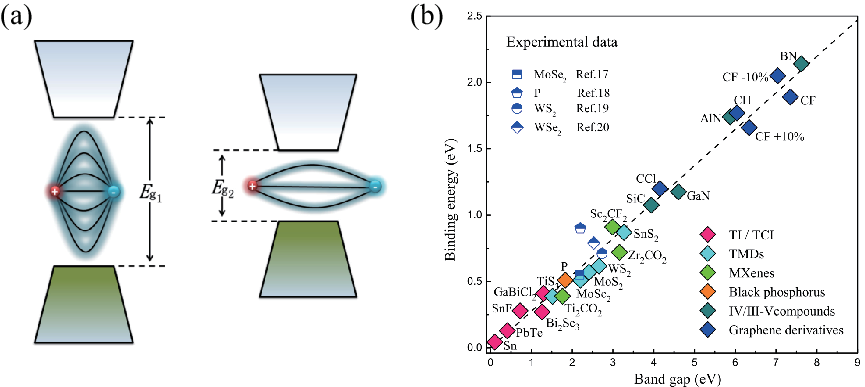}
\caption{\label{fig:fig1} (Color online) (a) A diagrammatic illustration of the correlation between $E_g$ and $E_b$. The larger the $E_g$, the weaker the screening effect, the stronger the exciton binding, the smaller the radius, and the higher the $E_b$. Conversely, the smaller the $E_g$, the stronger the screening, the weaker the exciton binding, the larger the radius, and the smaller the $E_b$. (b) The 1/4 linear scaling relationship between $E_b$ and $E_g$ in two-dimensional materials. Adapted from Ref. \onlinecite{Jiang2017}.}
\end{figure}

Being an EI requires $E_b > E_g$. Therefore, the larger $E_b$ is, the more favourable it is for EI formation. According to the hydrogen atom model, it can be seen that a decrease in dimension leads to an increase in $E_b$ and a decrease in exciton radius. For example, the ground state energy of a two-dimensional hydrogen atom is four times that of a three-dimensional one, and is even divergent in one dimension\cite{Zhao}. There is no doubt that low-dimensional materials offer more opportunities for realizing EI.

However, simply increasing $E_b$ is not enough, as $E_b$ and $E_g$ tend to show a subtle correlation. As illustrated in Fig. 1(a), decreasing $E_g$ makes electronic excitation less difficult, thus favouring exciton formation; at the same time, this leads to an increase in the system screening, which in turn inhibits the exciton formation. A zero-gap system is naturally more favourable for the EI formation, since any finite electron-hole interaction leads to the opening of an exciton gap. However, things are not so straightforward.

Let us consider the polarizability $\varepsilon$ of a system, which is calculated as
\begin{equation}\label{(1)}
\varepsilon = A\sum_{c,v} \int_{\rm \textbf{k}}\frac{|\langle \emph{u}_{\emph{c}, \rm \textbf{k}} |\nabla_{\rm \textbf{k}}| \emph{u}_{\emph{v}, \rm \textbf{k}}\rangle|^2}{\emph{E}_{\emph{c}, \rm \textbf{k}} - \emph{E}_{\emph{v}, \rm \textbf{k}}}d\rm \textbf{k}.
\end{equation}
Here $\emph{u}_{\emph{c}, \rm \textbf{k}}$ and $\emph{u}_{\emph{v}, \rm \textbf{k}}$ refer to the periodical parts of conduction and valence band Bloch states, respectively, and \textbf{k} is integrated over the first Brillouin zone. $A$ is a coefficient associated with the dimension. By definition, the minimum value of the denominator $\emph{E}_{\emph{c}, \rm \textbf{k}} - \emph{E}_{\emph{v}, \rm \textbf{k}}$ is $E_g$ of a direct-gap material. When $E_g\rightarrow 0$, one can find from Eq. (3) that, $\varepsilon$ tends to diverge. This means a zero, rather than a finite, $E_b$. Therefore, simply reducing $E_g$ or increasing $E_b$ does not guarantee the emergence of an EI\cite{Jiang2018}. Even in two-dimensional semiconductors, Choi et al.\cite{Choi} discovered a quantitative relationship between $E_b$ and $E_g$, which was subsequently refined by Jiang et al.\cite{Jiang2017} to a 1/4 linear scaling as shown in Fig. 1(b). However, the case becomes quite different when band-edge states $\emph{u}^{cbm}_{\emph{c}, \rm \textbf{k}}$ and $\emph{u}^{vbm}_{\emph{v}, \rm \textbf{k}}$ make the numerator $\langle \emph{u}^{cbm}_{\emph{c}, \rm \textbf{k}} |\nabla_{\rm \textbf{k}}| \emph{u}^{vbm}_{\emph{v}, \rm \textbf{k}}\rangle$ a constant 0. Now, the increase or decrease of $E_g$ has no effect on the integral of Eq. (3), that is, $\varepsilon$ does not change with the change of $E_g$. Since $\varepsilon$ characterizes the dielectric screening of the system, its invariance implies an invariant $E_b$. In other words, $E_g$ and $E_b$ are not correlated, but decoupled, and both become insensitive to changes in the other. At this point, if $E_g$ can be adjusted close to 0, EI will always emerge.

Within the first-principles BSE framework, $\langle \emph{u}_{\emph{c}, \rm \textbf{k}} |\nabla_{\rm \textbf{k}}| \emph{u}_{\emph{v}, \rm \textbf{k}}\rangle$ describes the dipole-moment. A non-zero dipole-moment means the formation of optically active bright excitons, while a near-zero dipole-moment means the formation of optically inactive dark excitons. We thus refer to $\langle \emph{u}^{cbm}_{\emph{c}, \rm \textbf{k}} |\nabla_{\rm \textbf{k}}| \emph{u}^{vbm}_{\emph{v}, \rm \textbf{k}}\rangle = 0$ as the dark-exciton rule, which states that the transitions between the involved band-edge states are optically forbidden. This is easy to understand. Dark excitons are not dissipated by radiative recombination and therefore have longer lifetimes. Such a dark-exciton rule explains why EIs are mostly present in indirect-gap semiconductors/semimetals and electron-hole bilayers. The former is due to momentum conservation while the latter is due to spatial separation of electrons and holes, both leading to $\langle \emph{u}^{cbm}_{\emph{c}, \rm \textbf{k}} |\nabla_{\rm \textbf{k}}| \emph{u}^{vbm}_{\emph{v}, \rm \textbf{k}}\rangle = 0$. Rather, this rule also suggests that excitonic instability is likely to occur in direct-gap systems\cite{Jiang2018}. Unlike the indirect-gap case, here the electron and hole are located at the same $k$-point in the Brillouin zone, and thus the Jahn-Teller intrusion such as that found in 1$T$-TiSe$_2$ will be fundamentally suppressed.

\vspace{0.3cm}
\textbf{3.2. EIs via band engineering}
\vspace{0.3cm}

There can be different ways to fulfil $\langle \emph{u}^{cbm}_{\emph{c}, \rm \textbf{k}} |\nabla_{\rm \textbf{k}}| \emph{u}^{vbm}_{\emph{v}, \rm \textbf{k}}\rangle$ = 0. For example, if the band-edge states have the same parity, transitions between them are dipole forbidden\cite{Jiang2018}. Two-dimensional III-V semiconductors in the double-layer honeycomb structure, such as AlSb, have their frontier states with the same odd parity\cite{Lucking,Ma,SunS,Dong2021}. First-principles BSE calculations indeed show that it possesses a negative $E_t$ (defined as $E_t = E_g - E_b$), indicative of a potential EI. It is noteworthy that such a two-dimensional AlSb has recently fabricated on graphene-covered SiC(0001) with a gap of 0.93 eV found\cite{Qin}.

%fig02
\begin{figure}[htbp]
\includegraphics[width=0.95\columnwidth]{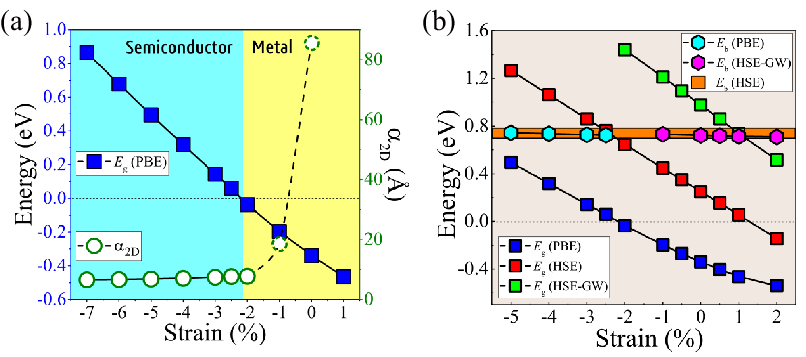}
\caption{\label{fig:fig2} (Color online) (a) Strain dependence of $E_g$ and two-dimensional polarizability $\alpha_{2D}$ for two-dimensional GaAs in the double-layer honeycomb structure. (b) Strain dependence of $E_g$ and $E_b$ with different calculation methods. Adapted from Ref. \onlinecite{Jiang2018}.}
\end{figure}

Equation (3) is more significant in that it shows that EIs can be made through band-engineering modulation of $E_g$ and/or $E_b$. The modulation means include strain, isoelectronic doping, and applied electric/magnetic fields, etc\cite{Jiang2018,Dong2020,ChenY,WangD}. A calculated dependence of $E_g$ and two-dimensional polarizability $\alpha_{2D}$ on strain for double-layer honeycomb GaAs is given in Fig. 2(a). It can be seen that $E_g$ is simply linear with respect to strain, while $\alpha_{2D}$ is almost unchanged in the positive $E_g$ region and diverges rapidly when the system becomes metallic. This confirms that $E_g$ is independent of the system screening. It should be particularly noted here that the strain-independent $\alpha_{2D}$ is not an inevitable consequence of the dark-exciton rule, but is determined by the unique band structure of GaAs\cite{Jiang2018}. Solving the BSE does find that $E_b$ is decoupled from $E_g$, as presented in Fig. 2(b). No matter which method is used, there is always a crossing point between the $E_b$ and $E_g$ curves. Therefore, although the present-day first-principles calculations are not yet able to predict $E_g$ accurately, the excitonic instability observed here is clearly independent of the computational method.

Similar physics exists in monolayer TiS$_3$. Unlike double-layer GaAs, its $E_g$ decreases with increasing compressive strain and a band reordering occurs near -2\%\cite{Jiang2018}. Fortunately, these bands in the vicinity of the Fermi energy have the same parity, so the band reordering does not lead to changes in the physics of interest. Compressive strain can equivalently be implemented by using isoelectronic substitutions of large radius atoms to generate a so-called chemical pressure. This was demonstrated experimentally by alloying Se into a monolayer of TiS$_3$\cite{Agarwal}. Calculations show that when Se replaces the middle S atoms, $E_g$ decreases to about half of TiS$_3$, whereas in contrast, $E_b$ keeps almost unchanged\cite{Dong2020}. As a result, in TiS$_2$Se, $E_g$ becomes smaller than $E_b$, i.e., Se doping drives the transition from band insulator to EI.

%fig03
\begin{figure}[htbp]
\includegraphics[width=0.9\columnwidth]{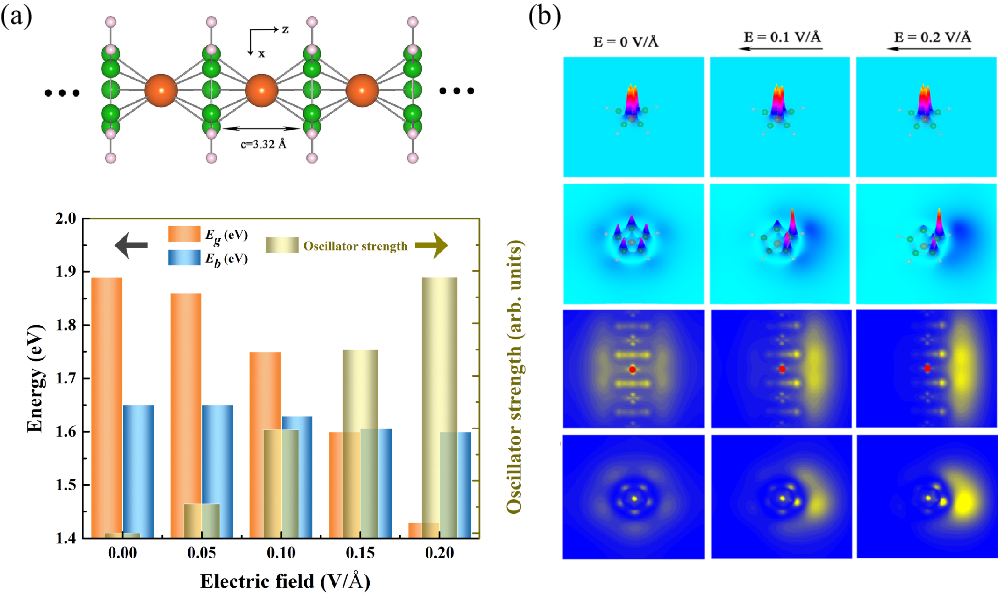}
\caption{\label{fig:fig3} (Color online) (a) Top row: Geometric structure of one-dimensional molecular wire MnCp$_\infty$. Bottom row: $E_g$, $E_b$ and oscillator strength of the lowest-energy exciton as a function of applied electric fields. On two sides of the critical electric field around 0.15 V/\AA, the relative magnitudes of $E_g$ and $E_b$ turn over, signalling the existence of a phase transition between the band insulator and the EI. It should be emphasised that the oscillator strengths in the EI phase from the current first-principles BSE have no physical significance, and the difference is only at the time of the phase transition. (b) Top two rows: Illustration of $E_g$ narrowing due to the giant Stark effect. The valence band top originates from the Mn $d$-orbitals. They are localized and are almost unaffected by the electric field, as shown in the first row. Therefore, the energy change of the valence band maximum due to the electric field is small. In contrast, the conduction band bottom originates from delocalized $\pi$-orbitals. As shown in the second row, these electrons move in the opposite direction to the external electric field and accumulate on one side of the two C atoms. The charge redistribution reduces the system's potential energy, compensating for the loss of kinetic energy, resulting in a significant downward shift of the conduction band. As a consequence, the $E_g$ is notably reduced. Bottom two rows: Evolution of the real-space wave function of the lowest-energy exciton under a transverse electric field. The red dots mark the positions of the holes (fixed at central Mn atoms). Adapted from Ref. \onlinecite{Liu2021}.}
\end{figure}

Applying an electric field can also drive an excitonic phase transition\cite{Liu2021}. As shown in Fig. 3(a), the $E_g$ of a one-dimensional molecular wire MnCp$_\infty$ decreases significantly with the enhancement of a transverse electric field. This is due to the fact that the frontier orbitals of the valence and conduction bands have different characteristics, and they respond differently to the electric field [see the top two rows of Fig. 3(b)]. The top of the valence band comes mainly from the $d$-orbitals of Mn and is less affected by the electric field. The bottom of the conduction band mainly comes from the $\pi$ orbitals formed by the $p$-electrons of the Cp ligands. Under an electric field, the delocalized $\pi$ electrons accumulate to one side, leading to a downward shift of the conduction band and a decrease in $E_g$. When the electric field is gradually increased to 0.2 V/\AA, the decrease of $E_g$ can reach 0.5 eV.

On the other hand, in the absence of applied electric field, the lowest-energy exciton of MnCp$_\infty$ is located at 0.24 eV, which is dark. Therefore, $E_b$ and $E_g$ are decoupled. This is confirmed by the calculations. As shown in Fig. 3(a), $E_b$ is very insensitive to the electric field. As the electric field is increased from 0 to 0.2 V/\AA, $E_b$ just decreases from 1.65 eV to 1.60 eV. The decrease of 0.05 eV is an order of magnitude smaller than that of $E_g$. Different dependence of $E_b$ and $E_g$ on the electric field results in the decreasing of $E_t$ with increasing electric field. When the field reaches about 0.15 V/\AA, $E_t$ starts to become negative, implying spontaneous exciton generation as an EI.

Another interesting aspect of electric field modulation is that it allows to change the optical activity of excitons\cite{Liu2021,Liu2022}. This may open up new opportunities for identifying the EI and building optical devices. We will talk about this later.

\vspace{0.3cm}
\textbf{3.3. Topological EIs}
\vspace{0.3cm}

%fig04
\begin{figure}[htbp]
\includegraphics[width=0.95\columnwidth]{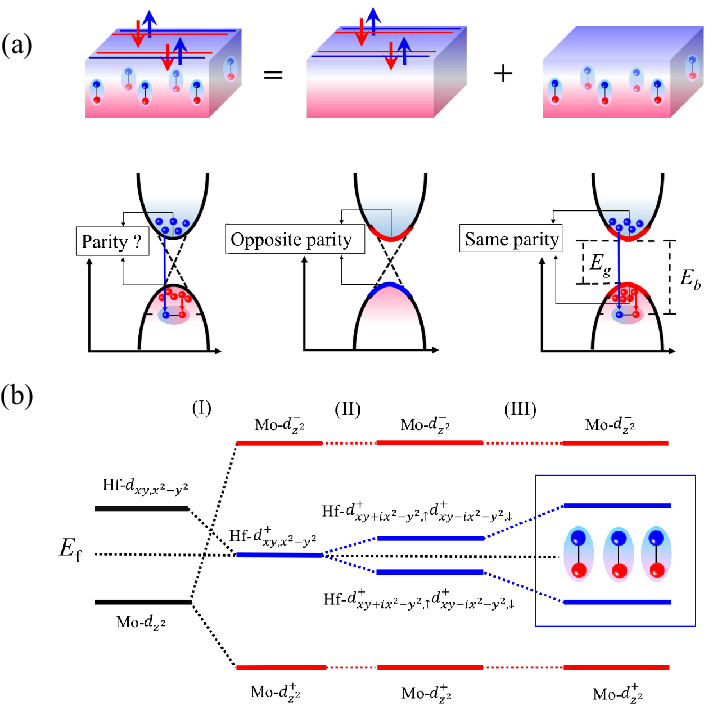}
\caption{\label{fig:fig4} (Color online) Schematics of (a) the parity frustration in a topological EI, and (b) the evolution from atomic orbitals into a topological EI of Mo$_2$HfC$_2$O$_2$ at the $\Gamma$ point. Adapted from Ref. \onlinecite{Dong2023}.}
\end{figure}

Topological EIs are a unique class of topological materials in which excitonic instability occurs. They may present new opportunities in the fields of both topological insulators and EIs\cite{Du,Dong2023,Dong2025,YangH,Varsano2020,Jia,Sun,Wang2019,HuL}. On the topological insulator side, firstly, spontaneous exciton condensation increases the bulk-gap, thus favouring higher operating temperatures. Secondly, the bulk-gap of topological EIs originates from excitons rather than spin-orbit coupling (SOC), which makes the material pool of topological EI not limited to heavy-element matter. Thirdly, exciton condensation depends on the overall screening effect of the system and therefore is not sensitive to the presence of some defects. On the EI side, although the charge neutrality of excitons prevents them from carrying electrical signals, topological edge states provide access for this purpose. In addition, the topological EI may also be an exotic fully supertransport system, i.e., exciton superfluidity in the bulk and dissipation-free transport at the edges.

Naturally, one wonders what kind of materials could be intrinsically topological EIs. Without loss of generality, let us consider a centrosymmetric system with a bandgap. Were it a topological insulator, it would require parity inversion. As a result, its band-edge states should have opposite parties. On the contrary, as pointed by the dark-exciton rule, excitonic instability occurs more often when the band-edge states have the same parity, so that the correlation between $E_g$ and $E_b$ is decoupled. This leads to the puzzle as illustrated in Fig. 4(a), i.e., should the band-edge states of topological EIs have the opposite or same parity? It is somewhat analogous to spin frustration, so called parity frustration\cite{Dong2023}.

The above analysis seems to suggest that intrinsic topological EI did not exist. As a matter of fact, one can understand the conventional SOC topological insulator as simply two events: (i) the existence of a SOC-opened bandgap; and (ii) the existence of state inversions with different parities. It is customary to put these two supposedly independent events together, thus creating a redundant constraint, i.e., the band-edge states undergo parity inversion. If parity inversion does not occur between band-edge states, it is possible to become topologically EI, e.g., double transition-metal MXenes V$_2$TiC$_2$F$_2$, Mo$_2$TiC$_2$O$_2$, Mo$_2$ZrC$_2$O$_2$ and Mo$_2$HfC$_2$O$_2$. First-principles BSE calculations show that their lowest excitons do have negative $E_t$\cite{Dong2023,YangH}.

The physics involved is explained in Fig. 4(b) using Mo$_2$HfC$_2$O$_2$ as an example. Its electronic structure evolution consists of three processes. In process (I), the 4$d_z^2$ orbitals of two Mo atoms hybridize to form bonding and antibonding states of opposite parities. At this point, the parity-inversion has occurred between the reconstructed Mo antibonding state and the Hf 5$d$ state. In process (II), turning on the SOC leads to the lift of double degeneracy at the Fermi energy, resulting in a topologically nontrivial gap. As can be seen, the SOC here is only responsible for gap opening, irrespective of the parity inversion. As a result, the band-edge states all come from Hf 5$d$ orbitals with the same parity. In process (III), the electron-hole interaction triggers an excitonic instability, resulting in a topological EI.

\vspace{0.3cm}
\textbf{3.4. Magnetic EIs}
\vspace{0.3cm}

%fig05
\begin{figure}[htbp]
\includegraphics[width=0.95\columnwidth]{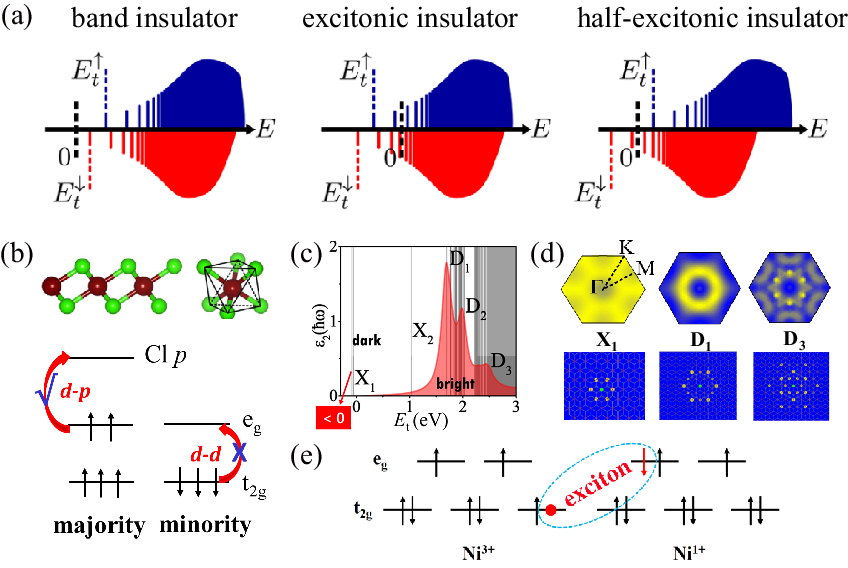}
\caption{\label{fig:fig5} (Color online) (a) A schematic representation of three kinds of excitonic instability in magnetic materials as classified by spin-resolved $E_t$. When both spins have $E_t > 0$, it is a normal magnetic insulator. When both spins have $E_t < 0$, it is a normal magnetic EI. When one spin has $E_t > 0$ and the other has $E_t < 0$, it is a new state of matter, i.e., half-EI. (b) Characterized $d$-electron configurations in monolayer 1$T$-NiCl$_2$. (c) The low-energy exciton energies (vertical lines), superimposed on the imaginary part of the BSE dielectric function. $X_i$ and $D_i$ denote dark and bright excitons, respectively. (d) Reciprocal-space (top) and real-space (bottom) exciton wave functions modulus. The green dots denote the hole positions that are fixed at the central Ni atom. (e) The spontaneous formation of $X_1$-excitons makes the neighbouring Ni atoms no longer equivalent, with one behaving like an electron and the other like a hole, as shown in (d). Thus, they exhibit +1 and +3 valence, respectively. The appearance of this mixed valence implies a change in the long-range magnetic interaction of the system from a super-exchange to a double-exchange mechanism. Adapted from Ref. \onlinecite{Jiang2019}.}
\end{figure}

It is well known that spontaneous exciton condensation can induce magnetism\cite{Zhitomirsky}, but whether excitonic instability occurs in intrinsically magnetic materials is not yet a question to be considered. There is no doubt that such excitonic instability gives rise to new physics as a result of the added spin degrees of freedom. Roughly speaking, excitonic instability in a spin-polarized system can be distinguished into three cases as shown in Fig. 5(a). (I) Both spins have positive $E_t$. This case corresponds to a conventional magnetic semiconductor. (II) Both spins have negative $E_t$, which corresponds to a magnetic EI. (III) One spin has a negative $E_t$ while the other has a positive $E_t$. This is a new state of matter analogous to a half-metal, which exhibits spontaneous exciton condensation in a single spin. It is named as the half-EI\cite{Jiang2019}.

First-principles BSE calculations identify monolayers of 1$T$-NiCl$_2$, NiBr$_2$, CoCl$_2$, and CoBr$_2$ as such half-EIs\cite{Jiang2019}. Noteworthy, monolayer CoCl$_2$ has been grown on highly oriented pyrolytic graphite substrates by molecular beam epitaxy\cite{CaiM}. Alternatively, they may be exfoliated from their layered bulk. Figure 5(b) interprets the relationship between the dark-exciton rule and the half-EI, taking NiCl$_2$ as an example. The Ni atom is situated in a local octahedral environment where its 3$d$-orbitals split into a threefold $t_{2g}$ and a twofold $e_g$. Its 8 $d$-electrons fully occupy the lower-lying $t_{2g}$ and spin-majority of higher-lying $e_g$. As a consequence, the majority spin features a $p-d$ gap, whose optical transition is dipole-allowed and forms a bright exciton. In contrast, the minority spin features a $d-d$ gap, whose optical transition is dipole-forbidden and forms a dark exciton. The dark-exciton rule predicts that the two spin channels may exhibit quite different excitonic instability behaviours. First-principles calculations show that only the $E_t$ of spin-minority is negative [see Fig. 5(c)], while the $E_t$ of spin-majority spin is positive. Thus, it is a half-EI.

Remarkably, here the excitonic instability emerges in a wide-gap semiconductor with a single-electron $E_g$ as high as 3.37 eV. This breaks with the conventional knowledge that EIs appear in semimetals or narrow-gap semiconductors\cite{Halperin}. Due to the giant $E_b$, the ground-state $X_1$-exciton is very localized and displays a clearly Frenkel feature, as reflected in Fig. 5(d). Its formation leads to two neighbouring Ni atoms with different valences, as illustrated in Fig. 5(e), and thus may affect the structural and magnetic properties. For example, unlike the super-exchange mechanism within the single-electron picture, the mixed valence of Ni is a hallmark of the double-exchange mechanism. In addition, the spontaneous exciton condensation of a single spin gives its two spins bosonic and fermionic behaviour, respectively, and thus very different responses to the external stimulus.

Monolayers of NiBr$_2$, CoCl$_2$ and CoBr$_2$ have similar physics. It is worth mentioning that the $E_t$ of CoCl$_2$ and CoBr$_2$ are as high as $-$0.59 and $-$0.54 eV, respectively. This implies that the excitonic instability may also be present in the fewer layers or even in their bulk parents. The main reason why there is such a large $E_t$ in the Co systems is that its $E_g$ comes from a further splitting of the Co $t_{2g}$ state, and is thus smaller than that of the corresponding Ni compounds\cite{Jiang2019}.

%fig06
\begin{figure}[htbp]
\includegraphics[width=0.95\columnwidth]{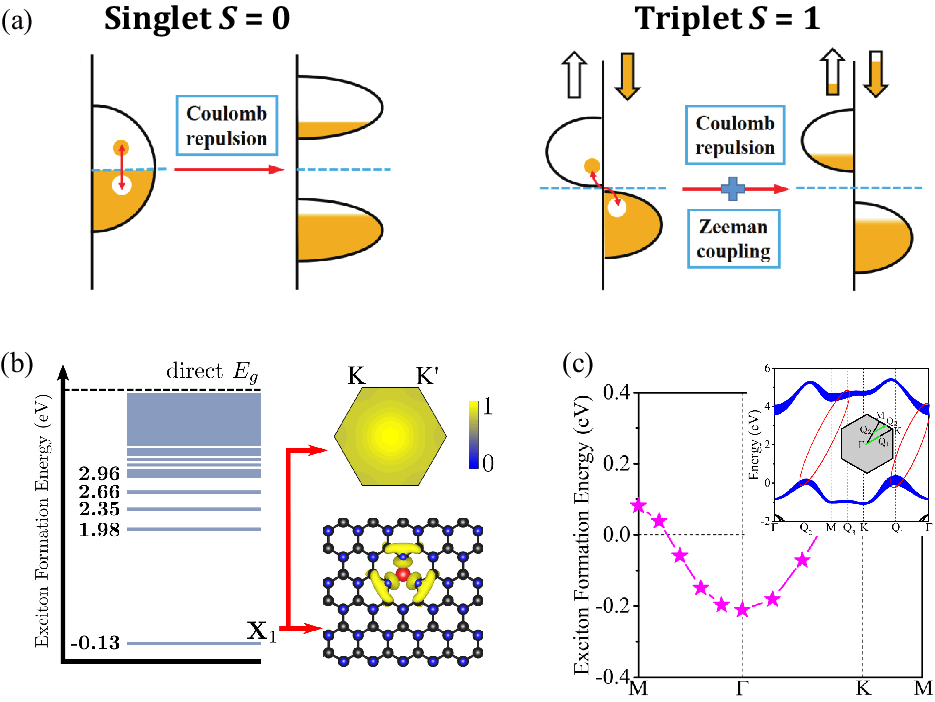}
\caption{\label{fig:fig6} (Color online) (a) A comparison of spin-singlet and spin-triplet EIs. When the electron transition occurs in the same spin, a singlet exciton with zero spin is formed. When there is a Zeeman split in the system and the electron transition into the opposite spin, a triplet exciton with spin 1 is formed. (b) First-principles BSE calculations reveal an excitonic instability in graphone with an $E_t = -$0.13 eV for the ground-state $X_1$-exciton. It is almost uniformly dispersed throughout the Brillouin zone and is very localized in real space. (c) Indirect-to-indirect characteristic transition between the single-electron band (inset) and the exciton band. The former has a $Q_1$$\Gamma$ indirect-gap, while the latter has a minimum energy of \textbf{q} = 0 excitons. The inset representatively demonstrates the physical essence of the non-local screening effect as the origin of the indirect-to-direct transition. Adapted from Ref. \onlinecite{Jiang2020}.}
\end{figure}

Another special feature of magnetic EI is embodied in the excitonic spin. Traditionally, excitons have always been associated with photoexcitation. Due to conservation of momentum, no spin-flip occurs when an electron is photoexcited. Therefore only singlet excitons with zero spin are considered [see the left panel of Fig. 6(a)]. However, in EIs where excitons form the ground state, the situation is quite different. All that is now concerned is whether the spontaneous production of excitons lowers the system energy, independent of light. Thus, the spin-selection rule provides us with another way to realize dark excitons.

Consider a spin-splitting system due to the Zeeman interaction as shown in the right panel of Fig. 6(a). If an excitonic instability occurs, the electrons and holes that make up this spontaneously generated exciton have the same spin. That is, it is a triplet exciton with spin 1, which will lead to a spin-triplet EI\cite{Jiang2020}. Since the triplet exciton carries spin, once these excitons condense to form a superfluid, a spin superfluidity is developed. The charge neutrality of the excitons prevents their condensation from generating super-transport of charge and thus cannot generate a superconduction-like signal. However, the condensation of triplet excitons can lead to spin superfluidity, thus providing new opportunities for experimental identification of EIs.

Single-side hydrogenated graphene (graphone) is predicted to be such a spin-triplet EI, as shown in Fig. 6(b)\cite{Jiang2020}. Here one of the two C $p_z$-orbitals forms a $\sigma$-bond with H. And the unsaturated $p_z$-orbitals undergoes a Zeeman-like splitting, with the electron occupying the lower-lying majority-spin band and emptying the higher-lying minority-spin band. Because the highest valence band and the lowest conduction band are both contributed by the same C atom, the ground-state excitons they make up are very localised in real space. As shown in Fig. 6(b), it extends only about 2 lattices. As a result, the exciton has a giant $E_b$ and exhibits at the same time a strong boson rather than electron-hole pair behaviour.

Real-space localization implies reciprocal-space delocalization, which leads to an unexpected indirect-to-direct crossover between single-electron and many-body bands as plotted in Fig. 6(c). Its single-electron band shows an indirect gap with the conduction band minimum at $\Gamma$ and the valence band maximum between $\Gamma$ and $K$. However, the calculated exciton band is direct, i.e., the \textbf{q} = 0 exciton has the lowest energy. This is virtually impossible in a three-dimensional system. In fact, it constitutes a direct and explicit evidence of non-local screening in the low-dimensional system. Since three-dimensional bulk materials have strong screening in all directions, one can simply describe it in terms of a single dielectric constant. The lowest-energy exciton at this point is usually localized only near the top of the valence band and the bottom of the conduction band in the reciprocal space. This is quite different in graphone, where the exciton is almost uniformly dispersed throughout the Brillouin zone, and the contributions of all $k$-points must be taken into account. For example, as can be seen in the inset of Fig. 6(c), in addition to the usual band-extrema transition between $\Gamma$$Q_1$, the $Q_2$$Q_3$ transition satisfies the same wavevector requirement. However, the latter corresponds to a single-electron gap of 4.6 eV, which is 1 eV larger than the 3.6 eV of the former.

In Eq. (2), the exchange interaction $V_{vckv'c'k'}$ of the BSE kernel controls the energy splitting between spin-singlet and spin-triplet excitons. Its repulsive nature raises the system energy, and thus the spin-triplet exciton is of lower energy. Based on this, spin-triplet EI is also predicted in a semiconducting diatomic Kagome lattice, e.g.,  a superatomic graphene\cite{Sethi}. Therein the large effective mass of the topologically flat bands drastically reduces the system screening and increases $E_b$. At the same time, the highly overlapping electron-hole wavefunctions enhance not only the Coulomb's direct interaction, but also the exchange interaction, which brings about singlet-triplet splitting up to 0.4 eV.

\vspace{0.3cm}
\textbf{4. Outlook and perspective}
\vspace{0.3cm}

%fig07
\begin{figure*}[htbp]
\includegraphics[width=0.95\columnwidth]{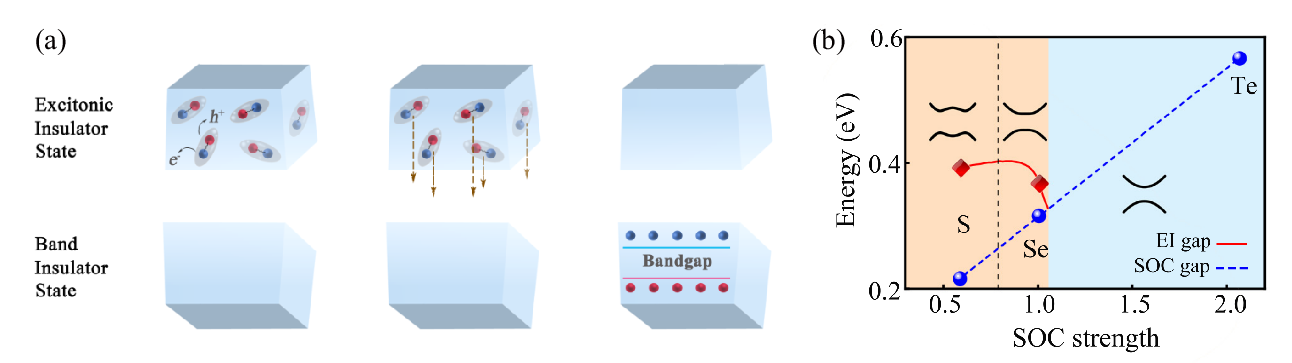}
\caption{\label{fig:fig7} (Color online) (a) A schematic diagram of the coherent radiation due to an external field-induced phase transition as a means of distinguishing between a ``bright" EI and a band insulator. Adapted from Ref. \onlinecite{Liu2021}. (b) Unique bulk-edge correspondence in LiFe$X$ ($X$ = S, Se, and Te) as a means of distinguishing between topological EIs and conventional SOC quantum anomalous Hall insulators. Briefly, since LiFeS and LiFeSe show excitonic instabilities while LiFeTe does not, LiFeSe has the smallest bulk-gap when all have dissipationless topological edge states. If the excitonic instability is not considered, they are all conventional SOC quantum anomalous Hall insulators. The bulk-gap at this point increases linearly from LiFeS to LiFeSe to LiFeTe (blue dashed line). Adapted from Ref. \onlinecite{QuarXiv}.}
\end{figure*}

Despite these advances, there are still many key issues to be resolved towards the first recognised EI. This involves the combined efforts and collaboration among the computational, theoretical, and experimental communities, including, but not limited to, searching for and designing suitable candidate materials, developing the theory of excitonic instability and spontaneous condensation, revealing the subtle correlation between spontaneous symmetry breaking and potential superfluidity, and exploring promising applications.

On the computational material design side, there is an urgent need to develop suitable and accurate methods for exciton calculations. Although first-principles BSE is widely used to solve $E_g$ and $E_b$, it has its limitations. On the one hand, the paradigm can only be applied to semiconductors. For semimetals with small overlap of valence and conduction bands, it remains an open question how to solve for the energetic structure of the exciton. This prevents a direct comparison of $E_g$ and $E_b$ to prove whether a semimetal-EI transition has occurred, e.g. in TiSe$_2$. The key is to develop a method for calculating the dielectric function in the presence of free charge carriers. On the other hand, by its theoretical definition, the EI is ultimately a matter of quantitative comparison between $E_g$ and $E_b$. However, current first-principles methods cannot accurately predict $E_g$. Underestimation of $E_g$ is a notorious problem of DFT-based electronic structure methods. Although Hubbard $U$, hybrid functional and GW can fix it to some extent, these methods are either not well compatible with subsequent BSE or too costly to calculate. It is thus very valuable to solve the $E_g$ problem at the DFT level. Perhaps AI techniques such as deep learning can play a role. For $E_b$, there is not even enough experience to qualitatively predict the performance of different methods, e.g., whether it is overestimated or underestimated. As a result, different customers may come to completely opposite conclusions because of their different methodological choices. For the same material, some may conclude that it is an EI with $E_t < 0$, while others may conclude that it is a non-EI with $E_t> 0$. Which method, at least statistically, is more reliable is a subject for further research.

On the theoretical side, many-body interactions remain a difficult problem in theoretical physics today. The formulation of BCS theory has greatly advanced the understanding of superconductivity and has also been borrowed to study EIs. The refinement of the EI theory urgently needs the facilitation of experiments, while on the other hand, experiments lack a powerful means to confirm the EI. This makes the theory and experiment unable to form complementary synergy and fall into the dilemma of ``chicken and egg".

While the dark-exciton rule is applicable to almost all known candidates, such as indirect-gap semimetals/semiconductors, quantum-well structures, wide-gap and even strongly-correlated semiconductors, it is neither a sufficient nor a necessary condition for the EI formation. Exploring the instability mechanism originating from optically active bright excitons is not only of fundamentally scientific significance, but also may provide new opportunities for EI identification\cite{QuAQT}. As depicted in Fig. 7(a), the excitons in an EI have the same phase due to the macroscopic quantum state nature. Whenever an EI-band insulator phase transition occurs, the excitons become unstable and electrons and holes start to recombine. If the excitons are optically active, the recombination emits coherent light. Such ``bright"-exciton EIs can then be identified by their quantum coherence properties. Therefore, although it does not matter whether it is a bright or a dark exciton at the time of the EI formation, there is a difference in the performance during the phase transition. Liu \emph{et al.}\cite{Liu2021} propose that an applied transverse electric field can simultaneously drive a band insulator-EI phase transition and activate the dark excitons by breaking the symmetry to realize a bright EI [see the lower two rows of Fig. 3(b)].

The lack of ``exclusive fingerprints" is arguably one of the most central bottlenecks in the field of EI today. There has been no answer to the question of what unique macroscopic phenomena are translated into by the microscopic mechanism of excitonic instability. At present, identifying an EI is usually done by using abrupt changes induced by exciton condensation reforming band edge states. However, in practice, such macroscopic changes may correspond to more than one microscopic mechanism, and it is difficult to exclude the involvement of all other possibilities. Whether a phenomenological definition of EI can be established, as in the case of superconductors, is at the heart of the matter. In this sense, the combination of EI with topological, magnetic and other orders may open new windows. For example, as shown in Fig. 7(b), a recent study suggests that the unique bulk-edge correspondence serves as an identifying fingerprint for topological EIs\cite{QuarXiv}. Although this does not ultimately solve the problem of a universal fingerprint for EIs, the interplay between different orderings has the potential to produce unique electrical or optical phenomena, which can be useful to first realize the experimental confirmation of a particular class of EIs.

This work was supported by the Ministry of Science and Technology of China (Grant Nos. 2023YFA1406400 and 2020YFA0308800), the National Natural Science Foundation of China (Grant No. 12474064).

\end{document}